\documentclass{article}

\usepackage{arxiv}
\usepackage[utf8]{inputenc} 
\usepackage[T1]{fontenc}    
\usepackage{hyperref}       
\usepackage{url}            
\usepackage{booktabs}       
\usepackage{amsfonts}       
\usepackage{nicefrac}       
\usepackage{microtype}      
\usepackage{lipsum}		
\usepackage{graphicx}
\usepackage{natbib}
\usepackage{doi}
\usepackage{listings}
\usepackage{xcolor}
\definecolor{codegreen}{rgb}{0,0.6,0}
\definecolor{codegray}{rgb}{0.5,0.5,0.5}
\definecolor{codepurple}{rgb}{0,0,205}
\definecolor{backcolour}{rgb}{0.95,0.95,0.92}
\lstdefinestyle{mystyle}{
    backgroundcolor=\color{backcolour},   
    commentstyle=\color{codegreen},
    keywordstyle=\color{magenta},
    numberstyle=\tiny\color{codegray},
    stringstyle=\color{codepurple},
    basicstyle=\ttfamily\footnotesize,
    breakatwhitespace=false,         
    breaklines=true,                 
    captionpos=b,                    
    keepspaces=true,                 
    numbers=left,                    
    numbersep=5pt,                  
    showspaces=false,                
    showstringspaces=false,
    showtabs=false,                  
    tabsize=2
}
\title{The openESEA Modelling Language for specifying Environmental, Social and Governance Accounting Methods}

\author{ \href{https://orcid.org/0000-0001-7343-4270}{\includegraphics[scale=0.06]{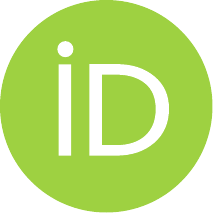}\hspace{1mm}Sergio España} \\
	Department of Information and Computing Sciences\\
	Utrecht University\\
	Utrecht 3584CC, the Netherlands \\
	VRAIN. Valencian Research Institute for Artificial Intelligence \\
	Universitat Politècnica de València \\
	46022 Valencia, Spain \\
	\texttt{s.espana@uu.nl} \\
	\And
    \href{https://orcid.org/0000-0002-3744-0013}{\includegraphics[scale=0.06]{orcid.pdf}\hspace{1mm}Vijanti Ramautar}\\
	Department of Information and Computing Sciences\\
	Utrecht University\\
	Utrecht 3584CC, the Netherlands \\
	\texttt{v.d.ramautar@uu.nl} \\
}

\renewcommand{\headeright}{Version 4}
\renewcommand{\undertitle}{Version 4}
\renewcommand{\shorttitle}{openESEA modelling language}

\hypersetup{
pdftitle={openESEA modelling language},
pdfsubject={ICS},
pdfauthor={Vijanti, Ramautar, Sergio, España},
pdfkeywords={Sustainability, Xtext, Accounting},
}

\begin{document}
\maketitle

\begin{abstract}
Over the years environmental, social and governance accounting (ESGA) has become a common practice among responsible organisations. ESGA entails assessing and reporting organisations" performance on environmental, social and governance topics. In this report, we present a textual grammar for specifying ESGA methods. With the grammar ESGA models can be created. Such models can be interpreted by our open-source, model-driven tool, called openESEA. The report presents the metamodel of the grammar, the grammar itself, and explanations of each grammar primitive.
\end{abstract}

\section{Introduction}
 Environmental, social and governance accounting (ESGA) is the process of assessing the social and environmental effects of an organisation’s actions, and reporting them to particular interest groups or to society at large~\cite{gray2000current,adams2007engaging}. In essence, it is the practice of assessing organisations' performance on governance, social, and environmental topics by calculating indicators. There are many ESGA methods, standards and frameworks that define how to perform ESGA~\cite{espana2023coperman}; for the sake of simplicity we will refer to them as ESGA methods. Examples of ESGA methods are ISO 14001:2015 standard \citep{ISO14001} by the International Organisation of \citep{ISO}, the GRI Standards \citep{GRI_Standards} by the Global Reporting Initiative\cite{GRI}, the UN Global Compact\citep{UNGC} by the United Nations Global Compact \citep{UN}, the Common Good Balance Sheet \citep{CGBS} by Economy for the Common Good \citep{ECG}, and the B Impact Assessment \citep{BIA} by B Lab \citep{BLab}. Many ESGA methods are supported by software tools. Most of these tools are rigid since can solely be used to assess the ESGA method they were developed for. To facilitate ESGA adoption and method tailoring, we are developing an open-source model-driven software tool. This tool can, in principle, support all ESGA methods that can be specified with the openESEA modelling language. Our tool is called openESEA~\cite{espana2019model,espana2022model}; it is open-source and web-based, and it is being valorised in the context of the Netherlands through the SCENTISS project\footnote{SCENTISS is a Dutch project funded by NWO, the Dutch Organisation for Scientific Research \url{https://scentiss.nl}}. Recently, the openESEA project has merged with the Show your Heart project, led by the Spanish Network of Networks of Alternative and Solidarity Economy\footnote{REAS - Red de Redes de Economía Alternativa y Solidaria \url{https://reas.red}} and the Catalan Network of Solidarity Economy\footnote{XES - Xarxa d'Economia Solidaria \url{https://xes.cat}}. This report focuses on describing the openESEA modelling language, which has been engineered as a domain-specific modelling language and has been subjected to several validations~\cite{ramautar2023csimq}.
 
 The openESEA tool works by interpreting textual models of ESGA methods. The models, for instance, contain general information of the method (e.g., name, description, version), a list of disclosure topics (e.g., gender equity, electricity consumption, working conditions), indicators (e.g., gender ratio, annual electricity consumption, minimum wage), surveys, and questions. All the model element are explained in Section~\ref{sec:explanation}. openESEA users can select an ESGA method and upload its model to openESEA. The tool parses the model and presents the model content in a user-friendly interface. openESEA can then be used to apply the selected method. 
 
 The textual models are specified according to a grammar. The grammar specifies the syntax of the model. For instance, the grammar states that every model should start with the name of the method by first writing \texttt{Name:}, followed by the name of the method. In the case of the B Impact Assessment, the first line of the model would be \texttt{Name:B Impact Assessment}. The full grammar can be found in Section~\ref{sec:grammar}

\section{Metamodel}

\begin{figure}[!ht]
    \centering
    \includegraphics[width=1\textwidth]{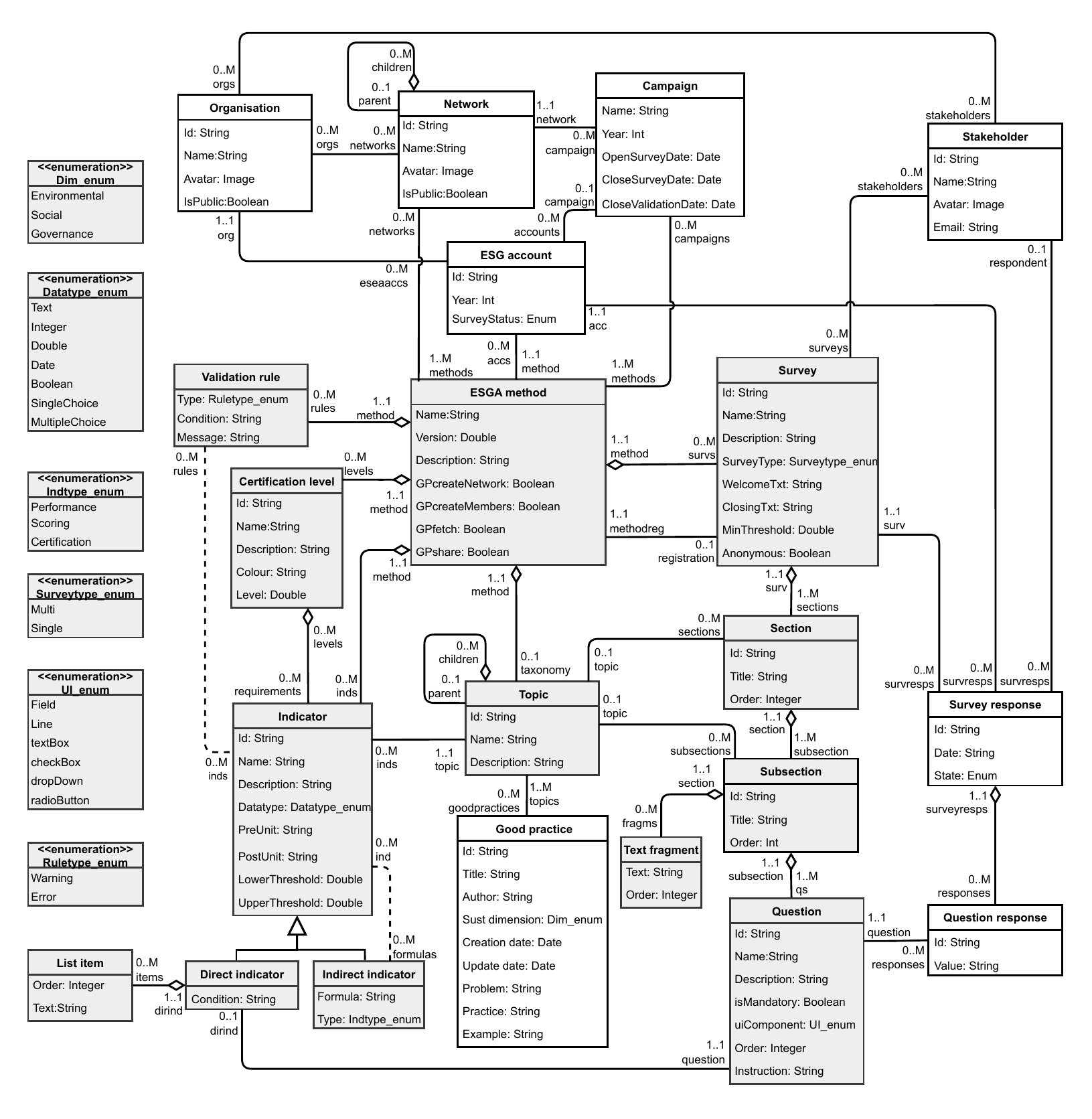}
    \caption{The openESEA metamodel V4}
    \label{fig:metamodel}
\end{figure}

To engineer the grammar we first created a metamodel, depicting all the grammar elements and their relationships. the first version of the openESEA metamodel was published in \cite{espana2019model}, but this report provides a more up-to-date version. In fact, this report is updated whenever a new version of the openESEA metamodel is available. Similarly as in \cite{espana2019model}, we opted for the language invention design pattern, in order to extend the metamodel. For our extension, we have analysed additional ESGA methods. We model ESGA methods, using the Process Deliverable Diagram (PDD) notation \citep{weerd2009meta}. The models are created based on method documentation and official information published by the organisation that developed the method. The models help us to better understand the ESGA process and the underlying data structure. The PDDs are validated with experts. After the validating interviews, the necessary modifications are done. The validated PDDs are used to analyse the ESGA methods. We do this by creating activity and concept comparisons. For this we adapt the method comparison approach~\cite{weerd2007developing}. The output of the method comparison is used to update the ESGA method metamodel. The metamodel is an UML Class Diagram\citep{OMG_UML}. Metrology standards informed our design decisions\citep{ISO99:2007}. The metamodel depicts the data structure of an ESGA method and its application. We differentiate between metaclasses that are generated when defining a method and metaclasses that are generated during the application of a method. Figure~\ref{fig:metamodel} depicts the metamodel. The metaclasses with a grey background are the metaclasses that are instantiated when specifying an ESGA method. Therefore, these classes are be part of the grammar. All metaclasses, including those with white background, are part of the database model of the openESEA interpreter.

\newpage

\section{Explanation of the Metaclasses}
\label{sec:explanation}
The following subsections provide a global overview of what each grammar primitive entails. 

\subsection{ESGA method} A specification of how ESGA should be performed, according to the creator(s) of the method. This includes the specifications of surveys, stakeholder groups, certification levels, topics, indicators, questions, but also guidelines on for example the reporting format or on whether the results of an accounting have to be published.

Examples of ESGA methods are the B Impact Assessment by B Labs, the Common Good Balance Sheet by Economy for the Common Good, the GRI Standards by the Global Reporting Initiative, UN Global Compact by the United Nations, etc.

Equivalent terms: social balance, social auditing, non-financial reporting, integrated reporting, environmental social and governance (ESG) reporting, sustainability reporting.

\subsubsection*{Attributes}
\begin{itemize}
    \item \textbf{Name.} This indicates which name a method is given by the creator.
    \item \textbf{Version.} This indicates the version of the method.
    \item \textbf{Description.} A short explanation of the method.
    \item \textbf{GPcreateNetwork.} It indicates whether staff of the organisation orchestrating the network that manages this method can create good practices.
    \item \textbf{GPcreateMembers.} It indicates whether staff of the organisations that are members of the network can create good practices.
    \item \textbf{GPfetch.} It indicates whether a network wants their members to see good practices of other networks.
    \item \textbf{GPshare.} It indicates whether a network wants to share their good practices with other networks.
\end{itemize}

\subsubsection*{Relationships}
\begin{itemize}
    \item \textbf{groups.} These are the groups of stakeholders as defined by the method.
    \item \textbf{survs.} The surveys that belong to the method.
    \item \textbf{taxonomy.} The taxonomy of topics that are defined in the method.
    \item \textbf{inds.}The indicators that need to be reported on.
    \item \textbf{levels}. The levels of certification that can be obtained when the method is applied accordingly.
    \item \textbf{accs.} These are the ESGA accounts that are produced over the years using this method.
    \item \textbf{networks.} The networks that prescribe the method.
    \item \textbf{campaigns.} The ESGA campaigns in which this method has been applied.
\end{itemize}

\subsection{ESGA account}
The state of the application of an ESGA method at a given moment in time. Organisations typically apply ESGA methods with a given frequency (e.g. every year an environmental, social and governance accounting is performed). Typically, it refers to the data provided by an organisation in a specific year.

\subsubsection*{Attributes}
\begin{itemize}
\item \textbf{Id.} This indicates which name a method is given by the creator.
\item \textbf{Year.} The year that an account belongs to. If the method defines \textit{campaigns}, then the year of the campaign should be copied to the year of the ESGA account.
\item \textbf{SurveyStatus.} Whether all surveys of an account are open or closed. \textit{Enumerated options\{open, closed\}}
\end{itemize}

\subsubsection*{Relationships}
\begin{itemize}
\item \textbf{org.} This is the organisation that has produced this account.
\item \textbf{surresps.} The survey responses belonging to that particular account.
\item \textbf{method.} The method according to which the ESGA account is performed. 
\end{itemize}

\subsection{Organisation}
A social entity that is goal-directed, is designed as a deliberately structured and coordinated activity system, and is linked to the external environment \cite{daft2015organization}. Within this research, we are interested in organisations that apply ESGA methods. 

Such organisations can have many different natures (for profit, vs non-profit), legal entity types (limited societies, associations, cooperatives) and belong to different sectors (e.g. utilities, information and communication technology, education, development aid, government). We assume them to be responsible organisations, who care about their organisational ethics and their social and environmental impacts. 

Equivalent terms: enterprise, company, entity.

\subsubsection*{Attributes}
\begin{itemize}
\item \textbf{Id.}  It is an identifier of the organisation, just to later use it for reference and links within the database. The user is not meant to edit this id.
\item \textbf{Name.}  It is the name of the organisation. E.g. HappyCows.
\item \textbf{Avatar.} It is an image that will be displayed in the tool when the organisation is presented. It could be a logo.
\item \textbf{isPublic}. Whether the organisation is shown to the general public or only to users with permissions to edit and see the organisation data.
\end{itemize}

\subsubsection*{Relationships}
\begin{itemize}
\item \textbf{stakeholders.} These are the \textit{stakeholder}s (i.e. people) that are related to the organisation.
\item \textbf{networks.} It indicates the \textit{network}s to which the organisation belongs.
\item \textbf{EseaAccs.} These are the \textit{ESGA account}s that the organisation has produced over the years.
\end{itemize}

\subsection{Network}
A group of responsible enterprises. Within this research we are interested in networks that offer or prescribe an ESGA method.

Often a network prescribes a specific ESGA method and all responsible enterprises within this network have to apply that method to become members (e.g. to be part of the B Corp network, members have to perform the B Impact Assessment). Other networks allow their members to select the ESGA method of their preference, or have a different means to determine the responsible nature of the organisations. Often, organisations need to demonstrate a certain performance in order to be granted membership; for instance, the ESGA method might have a scoring mechanism and the network defines a minimum threshold (e.g. B Impact Assessment produces a score from 0 to 200, and organisations need to score at least 80 to become B Corporations).

The network is typically managed (i.e. orchestrated) by an organisation, who also maintains (i.e. evolves and provides support to) the ESGA method. For instance, B Corporations is managed by B Labs. Sometimes the network and the organisation that manages it receive the same name (e.g. REAS Network of Networks of Alternative and Solidarity Economy, \textit{REAS Red de Redes de Economía Alternativa y Solidaria} in Spanish).

\subsubsection*{Attributes}
\begin{itemize}
\item \textbf{Id.} It is an identifier of the network, just to later use it for reference and links within the database. The user is not meant to edit this id.
\item \textbf{Name.} It is the name of the network. E.g. B Corporations, Economy for the Common Good, REAS Network of Networks of Alternative and Solidarity Economy.
\item \textbf{Avatar.} It is an image that will be displayed in the tool when the network is presented. It could be a logo.
\item \textbf{isPublic}. Whether the network is shown to the general public or only to users with permissions to edit and see the network data.
\end{itemize}

\subsubsection*{Relationships}
\begin{itemize}
\item \textbf{orgs.} This indicates which organisations are part of the network.
\item \textbf{method.} It indicates which methods should be applied in order to become part of the network.
\end{itemize}

\subsection{Stakeholder }
An individual with an interest or concern in something (e.g, one specific employee or one specific consumer).

\subsubsection*{Attributes}
\begin{itemize}
\item \textbf{Id.} It is an identifier of the stakeholder, just to later use it for reference and links within the database. The user is not meant to edit this id.
\item \textbf{Name.} It is the name of the stakeholder. E.g. "John Smith".
\item \textbf{Avatar.} It is an image that will be displayed in the tool when the stakeholder is presented. It could be a picture of the person.
\item \textbf{Email}. This is where the email address of the stakeholder is stored.
\end{itemize}

\subsubsection*{Relationships}
\begin{itemize}
\item \textbf{orgs.} This indicates which organisations are part of the network.
\item \textbf{survresps.} It indicates which methods should be applied in order to become part of the network.
\item \textbf{groups.} It indicates which methods should be applied in order to become part of the network.
\end{itemize}

\subsection{Survey} 
A survey is a questionnaire that a certain stakeholder group (e.g., workers, consumers, suppliers, etc.) has to respond in order to provide data for the direct indicators. Some surveys are meant to be responded by only one respondent (e.g., a manager), while other surveys are meant to be responded by several stakeholders of the same stakeholder group (e.g., all employees). We refer to the former as \textit{single-response} surveys and to the latter as \textit{multiple-response} surveys. But there is no explicit attribute to distinguish both types; actually, it is the number of instances of its corresponding stakeholder group which defines the type.

\subsubsection*{Attributes}
\begin{itemize}
    \item \textbf{Id.} It is an identifier of the survey, just to later use it for reference and links within the database. The user is not meant to edit this id.
    \item \textbf{Name.} It is the name of the survey E.g. “Quality of the working environment” could be a survey meant to be responded by all employees.
    \item \textbf{Description.} Brief explanation of the survey.
    \item \textbf{WelcomeTxt}. Text that is shown when the stakeholder opens the survey.
    \item \textbf{ClosingTxt}. Text that is shown when the stakeholder has finished the survey.
    \item \textbf{MinThreshold}.  Indicates the minimum number (or percentage) of responses the survey needs to receive for the outcome to be valid. For instance, this prevents that an employee satisfaction rate is calculated when only 2 employees out of 500 have responded the survey (thus preventing non-significant results or even fraud).
    \item \textbf{Deadline}. A due date by when the survey should have been responded and thus closes.  
    \item \textbf{SurveyType}. This indicates whether a survey is meant to be answered by a single respondent or by multiple. \textit{Enumeration options\{multiple; single\}}
\end{itemize}

\subsubsection*{Relationships}
\begin{itemize}
    \item \textbf{method.} The method the survey belongs to.
    \item \textbf{target.} This is the target stakeholder group; i.e. who has to respond this survey.
    \item \textbf{survresps.} These are the survey responses to the survey.
    \item \textbf{sects.} The sections that the survey has.
    \item \textbf{methodreg.} The registration survey belongs an ESGA method. This survey is single respondent and is only asked once.
\end{itemize}

\subsubsection*{Constraints}
\begin{itemize}
    \item The values of the attributes Section.Order should define a partial order, within the same survey; meaning that it is possible to omit some numbers but there should not be two sections of the same survey that share the same number. 
\end{itemize}

\subsection{Topic} 
\begin{figure}[!ht]
    \centering
    \includegraphics[scale=0.65]{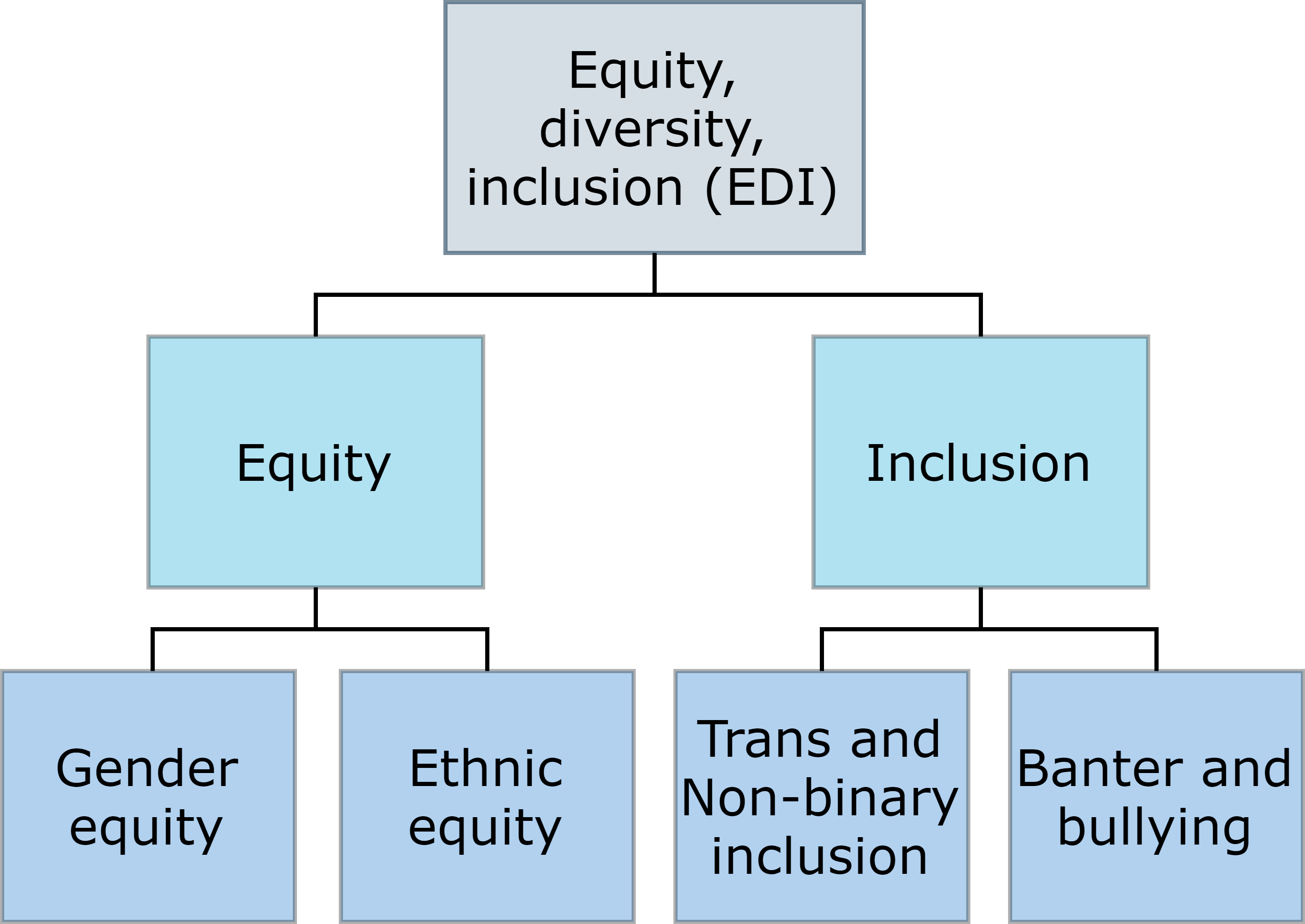}
    \caption{An example of a tree of topics}
    \label{fig:Tree_of_Topics}
\end{figure}

Topics group indicators concerning the same phenomenon together. For instance, “Gender equity” is a topic which groups all indicators concerning gender equity together (e.g., number of women in the staff, number of women in management positions, etc.). Another example is the topic “Environmental impact” which groups indicators concerning annual $CO_2$ emission, annual electricity consumption and annual waste together. Trees of topics can exist. This means that topics can be split up in more fine-grained topics. An example of a tree of topics is depicted in Fig.~\ref{fig:Tree_of_Topics}.

\subsubsection*{Attributes}
\begin{itemize}
    \item \textbf{Id.} It is an identifier of the topic, just to later use it for reference and links within the database. The user is not meant to edit this id.
    \item \textbf{Name.} The name of the topic. E.g. \texttt{``Gender equity''}
    \item \textbf{Description.} Brief explanation of the topic.
\end{itemize}

\subsubsection*{Relationships}
\begin{itemize}
    \item \textbf{children.} Subtopics of the parent topic.
    \item \textbf{inds.} These are the indicators that belong to a topic.
    \item \textbf{method.} The method that the topics are part of.
    \item \textbf{fragms.} The text fragments that explain or elaborate on the topics.
    \item \textbf{sections.} The sections that belong to a topic. 
\end{itemize}

\subsection{Text fragment} 
A text that explains or elaborates on the topic. 

\subsubsection*{Attributes}
\begin{itemize}
    \item \textbf{Id.} It is an identifier of the text fragment, just to later use it for reference and links within the database. The user is not meant to edit this id.
    \item \textbf{Text.} The topic-related text.
    \item \textbf{Order.} This indicates the order in which the text fragments should be displayed. 
\end{itemize}

\subsubsection*{Relationships}
\begin{itemize}
    \item \textbf{section.} The section to which the text fragment belongs to.
\end{itemize}

\subsubsection*{Constraints}
\begin{itemize}
    \item In the class \texttt{Question} there is an attribute called \texttt{Order} (i.e. \texttt{Question.order}). \texttt{Text fragment.Order} and \texttt{Question.Order} should not have the same value, since that would mean that the question and text fragment should be in the same place. 
\end{itemize}

\subsection{Section} 
A section groups subsections. Subsections group text fragments and questions in a survey.

\subsubsection*{Attributes}
\begin{itemize}
    \item \textbf{Id.} It is an identifier of the section, just to later use it for reference and links within the database. The user is not meant to edit this id.
    \item \textbf{Title.} The title is the name of the section
    \item \textbf{Order.} The order is an integer which indicates in which order the sections have to be displayed in the survey.
\end{itemize}

\subsubsection*{Relationships}
\begin{itemize}
    \item \textbf{surv.} The survey that the sections belong to.
    \item \textbf{subsections.} The subsections that compose the section.
    \item \textbf{topic.} This indicates which topic the section is about.
\end{itemize}

\subsection{Subsection} 
Subsections group text fragments and questions in a survey.

\subsubsection*{Attributes}
\begin{itemize}
    \item \textbf{Id.} It is an identifier of the subection, just to later use it for reference and links within the database. The user is not meant to edit this id.
    \item \textbf{Title.} The title is the name of the subsection
    \item \textbf{Order.} The order is an integer which indicates in which order the subsections have to be displayed in the survey.
\end{itemize}

\subsubsection*{Relationships}
\begin{itemize}
    \item \textbf{section.} The section that the subsection belongs to.
    \item \textbf{topic.} This indicates which topic the subsection is about.
    \item \textbf{fragms.} The text fragments that should be displayed in a particular subsection.
    \item \textbf{qs.} The questions that belong to a subsection.
\end{itemize}

\subsection{Indicator} An indicator is the definition of a measure that is assessed and reported on during the accounting. For instance, an indicator for the topic \texttt{``Gender equity''} could be the \texttt{``Gender pay gap''}. Indicators are composed of direct and indirect indicators. \texttt{``Gender pay gap''} is an indirect indicator, while the \texttt{``Average salary for men''} and \texttt{``Average salary for women''} are direct indicators.

\subsubsection*{Attributes}
\begin{itemize}
    \item \textbf{Id.} It is an identifier of the indicator, just to later use it for reference and links within the database. The user is not meant to edit this id.
    \item \textbf{Name.} Name of the indicator. E.g. \texttt{``Average salary for men''}
    \item \textbf{Description.} A short description of the indicator
    \item \textbf{Datatype.} The data type of the values expected for this indicator. \textit{Enumerated options\{text; double; date; boolean; singlechoice; multiplechoice\}}
    \item \textbf{PreUnit.} A unit that is placed as a prefix when displaying the value of this indicator. E.g. “€” as in a value of “12.50” meant to be displayed as “€ 12.50.” Other examples are “€”, ``\$''
    \item \textbf{PostUnit.}  A unit that is placed as a suffix when displaying the value of this indicator. E.g. ``Tons of CO2 equivalents'' as in a value of ``600'' meant to be displayed as ``600 Tons of CO2 equivalents''. Other examples are ``kg'', ``m2'', ``°C'', ``days'', ``MMBtu'', ``points'', ``KWh''
    \item \textbf{LowerThreshold.} A left boundary limit that is defined for a reasonable range. A value in the indicator below this limit should be considered suspicious; that is, a possible error (for instance, a yearly salary lower than the minimum wage). If not inserted, the anomaly detection algorithm that is part of the auditing feature will calculate 3 times the standard deviation (this is a common empirical standard).
    \item \textbf{UpperThreshold.} A right boundary limit that is defined for a reasonable range. A value in the indicator above this limit should be considered suspicious; that is, a possible error (for instance, a yearly salary four times the minimum wage is odd within organisations of the social and solidarity economy). If not inserted, anomaly detection will calculate 3 times standard deviation.
\end{itemize}

\subsubsection*{Relationships}
\begin{itemize}
    \item \textbf{method.} This shows which ESGA method the indicator belongs to.
    \item \textbf{topic.} This shows which topic the indicators are related to.
    \item \textbf{formulas.} The set of  indirect indicators that refer to this indicator in their formulas.
    \item \textbf{levels.} This indicates which certification levels can be obtained.
\end{itemize}

\subsubsection*{Constraints}
\begin{itemize}
    \item The relationship items is empty unless the indicator datatype is list.
\end{itemize}

\subsection{Direct indicator} 
The value of a direct indicator can be provided by a stakeholder via a question in a survey. Therefore a direct indicator does not have a formula. The direct indicator is a specialisation of the class \texttt{``Indicator''}, so it inherits all the attributes and relationships. At the moment, the value for direct indicators can only be collected through surveys; however, in the future, we envision that the values can also be directly queried from a database (e.g. obtaining the number of organisational members from the human resource management information system) or from sensors or smart devices (e.g. obtaining a measurement of water quality level from a sensor, or the energy consumption from a smart meter). This is why ``Direct indicator'' and ''Question'' are decoupled.

\subsubsection*{Attributes}
\begin{itemize}
    \item \textbf{Help.} A text that explains how to provide a value for this indicator.
    \item \textbf{Condition.} A logical formula (i.e. calculating a Boolean value) that determines whether the direct indicator is applicable to a given organisation. The formula can use references to other indicators. Optional.
\end{itemize}

\subsubsection*{Relationships}
\begin{itemize}
    \item \textbf{qs.} This shows which question asks a stakeholder for the value of the direct indicator.
    \item \textbf{items.} This indicates the set of items in a list that are available, in case the indicator datatype is singlechoice or multiplechoice.
\end{itemize}

\subsubsection*{Constraints}
\begin{itemize}
    \item With regard to conditions, circular references should be avoided. E.g. the condition should not refer to the same direct indicator where the condition is being defined. Similarly, the conditions of two different direct indicators should not refer to each other.
    \item It is only possible to create links of the type items when the indicator Datatype is \textit{singlechoice} or \textit{multiplechoice}.
\end{itemize}

\subsection{List item} 
If the datatype of a direct indicator is \textit{singlechoice} or \textit{multiplechoice} (for example in a question asking for the industry sector), then this class defines the answer options available in that list.

\subsubsection*{Attributes}
\begin{itemize}
    \item \textbf{Order.} An integer meant to specify the order in which the list items are presented to the survey respondent.
    \item \textbf{Text.} The label of the item. E.g. \texttt{``Yes''} or \texttt{``No''}; or the names of the industry sectors.
\end{itemize}

\subsubsection*{Relationships}
\begin{itemize}
    \item \textbf{dirind.} This indicates the direct indicator that the list item is part of.
\end{itemize}

\subsubsection*{Constraints}
\begin{itemize}
    \item Two list items of the same direct indicator cannot have the same text.
\end{itemize}

\subsection{Indirect indicator}  
An indirect indicator has a formula. An indirect indicator is calculated by using direct or other indirect indicators. The indirect indicator class can be used to define scoring rules, by creating a scoring indicator and defining a formula. The indirect indicator is a specialisation of the class ``Indicator'', so it inherits all the attributes and relationships. The indicator grammar primitive allows the method engineer to state whether the indicator type is ``Direct'' or ``Indirect''. In case of an indirect indicator, the method engineer can define a formula. There are several rules that ensure the syntax of the formula can be parsed by openESEA. These rules start in the grammar at line 91 until line 140.

\subsubsection*{Attributes}
\begin{itemize}
    \item \textbf{Formula.} The formula shows how an indirect indicator is calculated.
    \item \textbf{Type.} It defines whether the indicator is meant for assessing organisational performance (in terms that are meaningful to the stakeholders; e.g. ``gender\_ratio'') or whether it is related to scoring the organisation (probably with the intention to later check whether the organisation achieves a certification level; e.g. ``gender\_ratio\_score''). \textit{Enumerated options\{performance; score\}}
\end{itemize}

\subsubsection*{Relationships}
\begin{itemize}
    \item \textbf{ind.} This is actually a derived relationship. The formula defines which other indicators are used to calculate the value of the indirect indicator (we refer to such indicators as operand indicators because they are used in mathematical operations to calculate the indirect indicator). However, the formula is a text string. The ind relationship makes the links to the operand indicators explicit in the database.
\end{itemize}

\subsubsection*{Constraints}
\begin{itemize}
    \item With regard to formulas, circular references should be avoided. E.g. the formula should not refer to the same indirect indicator where the formula is being defined. Similarly, the formulas of two different indirect indicators should not refer to each other.
\end{itemize}

\subsubsection*{Examples}
\begin{itemize}
    \item Regular indirect indicators:
    \begin{itemize}
        \item \texttt{gender\_ratio: formula = women\_staff / (total\_staff)}
    \end{itemize}
    \item Scoring indirect indicators:
    \begin{itemize}
        \item \texttt{gender\_ratio\_score: formula = ``IF gender\_ratio < 0.15 THEN gender\_ratio\_score = 0 ELSE IF gender\_ratio < 0.85 THEN gender\_ratio\_score = 10 * (gender\_ratio - 0.15 ) / (0.85 - 0.15) ELSE gender\_ratio\_score = 10''}
    \end{itemize}
\end{itemize}

\subsection{Validation rule}  
A validation rule defines a logical condition over the values of one or more direct indicators. E.g. \texttt{NOT(F10="yes" AND F11="yes")}.The intention is to check whether the data in a survey response is correct. When an error or a warning are detected, then the user is prompted a message. See more examples below.

\subsubsection*{Attributes}
\begin{itemize}
    \item \textbf{Type.} Indicates the type of rule; that is, whether the rule is intended to detect errors (i.e. if the rule condition resolves to False, then there is an error in one or more question responses that needs to be corrected) in the data or whether it is intended to raise warnings about data that could be wrong (i.e. if the rule condition resolves to False, then the tool should invite the user to revise the data in one or more question responses). \textit{Enumerated options{Error; Warning}}
    \item \textbf{Condition.} It is a condition over the possible values of one or many direct indicators. This is an expression that resolves to a Boolean value. It is expected to be True; otherwise, the warning or error message should be shown to the user. 
    \begin{itemize}
        \item Note that only direct indicators can be used in the condition and not indirect indicators. This is something we might need to change later on, if we want to use validation rules for aggregation or auditing purposes.
    \end{itemize}
    \item \textbf{Message.}  message that is shown to the user once an error or warning is detected through a validation rule.    
\end{itemize}

\subsubsection*{Relationships}
\begin{itemize}
    \item \textbf{method.} The method that the validation rule is part of.
    \item \textbf{inds.} The direct indicators of which the validation rule validates the value. This relationship is not implemented in the grammar because it is redundant with the value of the attribute Condition (it refers to the direct indicators that are mentioned in the condition). It is, therefore, a derived relationship.
\end{itemize}

\subsubsection{Error examples}
\begin{itemize}
    \item STARS.
    \begin{itemize}
        \item \texttt{NOT(F10="yes" AND F11="yes")}
        \item \texttt{NOT(F11="no" AND F10="no" AND F12= "yes")}
    \end{itemize}
    \item XES Social Balance.
    \begin{itemize}
        \item Personnel expenses (q1206) vs salaries (q1102 + q1107)
        \begin{itemize}
            \item If salaries> 0, personnel expenses cannot be 0
            \begin{itemize}
                \item \texttt{Type: error}
                \item \texttt{Condition: (q1102+q1107) > 0 AND q1206 > 0}
                \item \texttt{Message: “When there are salaries  (q1102, q1107), the personnel expenses (q1206) cannot be 0”}
            \end{itemize}
            \item If salaries = 0, personnel expenses cannot be > 0
            \begin{itemize}
                \item \texttt{Type: error}
                \item \texttt{Condition: (q1102+q1107) = 0 AND q1206 = 0}
                \item \texttt{Message: “There are no salaries (q1102, q1107), so the personnel expenses (q1206) must be zero”}
            \end{itemize}
            \item Personnel costs must be higher than salaries, at least 15\% more.
            \begin{itemize}
                \item \texttt{Type: error}
                \item \texttt{Condition: q1206 >= 0.15 * (q1102+q1107)}
                \item \texttt{Message: “Personnel costs (q1206) must be higher than salaries (q1102+q1107) , at least 15\% more.”}
            \end{itemize}
        \end{itemize}
    \end{itemize}
\end{itemize}

\subsubsection{Warning examples}
\begin{itemize}
    \item Raise a concern when the average hourly salary is too low (e.g. below the lowest minimum salary in the Netherlands):
    \begin{itemize}
        \item \texttt{Type: warning}
        \item \texttt{TCondition = average\_hourly\_salary >=  10.77}
        \item \texttt{TMessage = “Make sure you entered a correct average hourly salary, since it seems quite low”)}
    \end{itemize}
    \item Raise a concern when the average hourly salary is too high (e.g. below the lowest minimum salary in the Netherlands):
    \begin{itemize}
        \item \texttt{TType: warning}
        \item \texttt{TCondition = average\_hourly\_salary <= 500} 
        \item \texttt{TMessage = “Make sure you entered a correct average hourly salary, since it seems quite high”}
    \end{itemize}
\end{itemize}

The rationale for these rules are that (i) the hourly minimum wage as of 1 July 2020 in The Netherlands, for 36 hours per week of a person of 21 or older, is € 10.77, so it could be that the user made a mistake with the decimal comma, and (ii) the minimum wage as of 1 July 2020, per month, for a kid of 15 years old, is 504€, so it could be that the user is entering monthly or yearly salaries, instead of hourly ones (\href{https://www.government.nl/topics/minimum-wage}{link} to the definition of minimum wages in the Netherlands).

\subsubsection*{Examples}
\begin{itemize}
    \item Regular indirect indicators:
    \begin{itemize}
        \item \texttt{gender\_ratio: formula = women\_staff / (total\_staff)}
    \end{itemize}
    \item Scoring indirect indicators:
    \begin{itemize}
        \item \texttt{gender\_ratio\_score: formula = “IF gender\_ratio < 0.15 THEN gender\_ratio\_score = 0 ELSE IF gender\_ratio < 0.85 THEN gender\_ratio\_score = 10* (gender\_ratio - 0.15 ) / (0.85 - 0.15) ELSE gender\_ratio\_score = 10”} 
    \end{itemize}
\end{itemize}

\subsection{Question} 
A question asks for the value of a direct indicator. 

\subsubsection*{Attributes}
\begin{itemize}
    \item \textbf{Id.} It is an identifier of the question, just to later use it for reference and links within the database. The user is not meant to edit this id.
    \item \textbf{Name.} The question sentence. E.g. \texttt{``What is the average salary of the women in your organisation?''}
    \item \textbf{Description.} A short description of the question. It can include help to answer properly, or examples.
    \item \textbf{IsMandatory.} Whether the question is obligatory or not.
    \item \textbf{UIcomponent.} Whether the question is an open question, multiple choice, etc. \textit{Enumerated options\{field; line; textbox; checkbox; radiobutton; dropdown\}}
    \item \textbf{Order.} An integer value meant to specify the order in which questions and text fragments are presented to the user in a survey section.
\end{itemize}

\subsubsection*{Relationships}
\begin{itemize}
\item \textbf{responses.} This shows the question response to the question.
\item \textbf{subsection.} This shows which section of the survey the question belongs to.
\item \textbf{dirind.} This indicates which direct indicator value the question requests.
\end{itemize}

\subsubsection*{Constraints}
\begin{itemize} 
    \item The values of the attributes Question. Order should define a partial order for each section; meaning that it is possible to omit some numbers but there should not be two sections of the same section that share the same number. 
    \item There are constraints to the possible user interface components that each type of question allows, as well as to the statistical functions that can be applied to the responses
\end{itemize}

\subsection{Survey response} 
The response of a survey made by a stakeholder. It contains many question responses.

\subsubsection*{Attributes}
\begin{itemize}
    \item \textbf{Id.} It is an identifier of the survey response. The user is not meant to edit this id.
    \item \textbf{Date} The timestamp when the stakeholder responded to the survey.
    \item \textbf{State.} An indication of whether the stakeholder is still editing the response (i.e. incomplete) or it is already complete. Only complete questions should be considered valid data. It has an enumerated data type with possible values ``incomplete'' and ``complete''. \textit{Enumerated options\{incomplete; complete\}}
\end{itemize}

\subsubsection*{Relationships}
\begin{itemize}
    \item \textbf{surv.} The survey the stakeholder has responded to respondents.
    \item \textbf{respondent.} The stakeholder that has filled in this response. While conceptually, we might argue that the respondent of a ``Survey response'' is known, in the an ESGA tool it is better to guarantee the anonymity of respondents of multiple-stakeholder surveys. While the tool should offer the possibility to send invitations to fill in the survey to all members of a stakeholder group (e.g. all employees of a company), no user of the tool should be able to trace back a response to the respondent. Only the tool should keep thattrace, for auditing purposes. Also, no user of the tool should be able to consult individual responses of a multiple-stakeholder survey, since it could be possible to identify a respondent indirectly through a combination of responses (it could be that there is only one racialised member in an organisation who identifies with a non binary gender), but they should just have access to aggregations of the overall responses to the survey (e.g. average, min, max, sum). 
    \item \textbf{acc.} The account that this response belongs to.
    \item \textbf{responses.} The set of responses for each question in the survey.
\end{itemize}

\subsubsection*{Constraints}
\begin{itemize}
    \item The year of the date of survey response should be the same as the year in ESGA account.
    \item A stakeholder should not respond more than once to the same survey for the same account.
\end{itemize}

\subsection{Question response} 
The response to a question is stored in this class.

\subsubsection*{Attributes}
\begin{itemize}
    \item \textbf{Id.} It is an identifier of the question, just to later use it for reference and links within the database. The user is not meant to edit this id.
    \item \textbf{Value.} The response to a question.
\end{itemize}

\subsubsection*{Relationships}
\begin{itemize}
    \item \textbf{question.} This indicates to which question the question response belongs to.
    \item \textbf{surveyresps.} This indicates to which survey the question response belongs to.
\end{itemize}

\subsection{Certification level} 
A certification is an official document attesting to a status or level of achievement. The certification  is issued by the network once the applicable requirements are met (note: The requirements can be specified as an indirect indicator).

\subsubsection*{Attributes}
\begin{itemize}
    \item \textbf{Name.} The name of the certification level.
    \item \textbf{Description.} The explanation of what the certification levels means.
    \item \textbf{Colour.} The colour that corresponds to the certification level, for display in the user interface. E.g. \texttt{``Bronze''}
    \item \textbf{Level.} A number representing the certification level. When there are several certifications within the same ESGA method, this number clarifies their order. E.g. \texttt{``3''}
\end{itemize}

\subsubsection*{Relationships}
\begin{itemize}
    \item \textbf{reqs.} The requirements to obtain a certification. It is a set of indicators whose formulas compute boolean values. A certification is achieved when all of them are \texttt{``True''}.
    \item \textbf{method.} The method that defines this certification.
\end{itemize}

\subsection{Good practice} 
Procedures that are accepted or prescribed as being correct or most effective. Examples of good practices in the ESGA domain are “only purchase eco-labelled products”, “use the most advanced and energy-optimised technologies and devices” or “develop a process for the continuous measurement of energy”.

\subsubsection*{Attributes}
\begin{itemize}
    \item \textbf{Id.} It is an identifier of the good practice, just to later use it for reference and links within the database. The user is not meant to edit this id.
    \item \textbf{Title.} A descriptive name for the good practice, in a single sentence.
    \item \textbf{Author.} The person who created the good practice.
    \item \textbf{Sust dimension.} A name of the coarse-grained topic that better relates to this good practice. \textit{Enumerated options\{Environmental; Social; Governance\}} 
    \item \textbf{Creation date.} The date on which the good practice is created.
    \item \textbf{Update date.} The date on which the good practice has been last updated.
    \item \textbf{Problem.} The problem(s) that the good practice helps alleviate or solve.
    \item \textbf{Practice}. The description of the good practice; i.e. the solution that an organisation can implement in order to solve the problem.
    \item \textbf{Example}. If necessary and available, an example clarifies how a specific organisation has applied the good practice.
\end{itemize}

\subsubsection*{Relationships}
\begin{itemize}
     \item \textbf{topics.} The good practice solves or alleviates a problem that is related to a topic.
\end{itemize}

\subsection{Campaign} 
A period during which the ESGA accounting can be performed. Examples of ESGA methods that work with campaigns are XES Social Balance and CDP.

\subsubsection*{Attributes}
\begin{itemize}
    \item \textbf{Name.} The name of the campaign, for instance the year.
    \item \textbf{OpenSurveyDate.} The date on which all surveys that belong to the method open.
    \item \textbf{CloseSurveyDate.} The date on which all surveys that belong to the method close.
    \item \textbf{CloseValidationDate} The date on which the validation of the account data should be finished. Auditors cannot ask organisations to adjust data after this date.
    \item \textbf{Year.} The year for which a campaign collects data.
\end{itemize}

\subsubsection*{Relationships}
\begin{itemize}
    \item \textbf{network.} The network that starts and ends the campaign. Every network can specify their own campaign.
    \item \textbf{method.} The ESGA method that the campaign belongs to. 
\end{itemize}

\subsubsection*{Constraints}
\begin{itemize}
    \item When the network administrator has defined a campaign, then no organisation should create more than one account per campaign.
\end{itemize}

\section{ESGA Grammar}
\label{sec:grammar}
After creating the metamodel we engineered a textual Xtext grammar based on the metamodel. The transformation from metamodel to textual grammar is done manually. The grammar can be used with an editor with syntax highlighting, completion suggestions, and warning and error messages. We chose to transform the metamodel into a textual grammar since initially, it requires less effort to produce a functioning editor for textual grammars. Moreover, using metamodels and grammars together may enhance tool interoperability \citep{neubauer2019reusable}. To create the grammar we use the Xtext framework \citep{Xtext}. Xtext is developed in the Eclipse Project \citep{Eclipse_Project} as part of the Eclipse Modeling Framework Project \citep{Eclipse_Modeling_Framework} and is licensed under the Eclipse Public License \citep{Eclipse_Public_License}. 

\definecolor{xtextblue}{RGB}{0,101,189}
\definecolor{xtextorange}{RGB}{227,114,34}
\definecolor{xtextolive}{RGB}{162,173,0}
\definecolor{xtextgray}{gray}{0.95}
\lstdefinelanguage{Xtext}{
	keywords=[1]{ID, INT, STRING, terminal, current, returns, grammar, with, generate, import, hidden, as, enum},
	keywordstyle=[1]\color{gray},
	keywords=[2]{name},
	keywordstyle=[2]\color{xtextorange},
	morecomment=[l]{//}, 
	morecomment=[s]{/*}{*/}, 
	morestring=[b]",
	morestring=[b]',
	moredelim=**[is][\color{gray}]{`}{`},
	stringstyle=\color{xtextblue}\ttfamily,
	tabsize=4
}

    
\begin{lstlisting}[language=Xtext, caption= ESGA grammar V3, backgroundcolor = \color{xtextgray}]
ESEA_method: 
	"Name:" STRING 
	"Version:" DOUBLE
	"isPublic:" BOOLEAN
	"Description:" STRING	
    "GPcreateNetwork:" BOOLEAN
    "GPcreateMembers:" BOOLEAN

	list_of_topics+=List_of_topics
	list_of_indicators+=List_of_indicators
	list_of_surveys+=List_of_surveys
	(list_of_certification_levels+=List_of_certification_levels)?
	(list_of_validation_rules+=List_of_validation_rules)?
	(registration_survey=Survey)?
	//Constraint: The registration survey should be single respondent
;

List_of_topics:
	// Constraint: 1 and only 1 root topic
    "Topics:"
	(topic+=Topic)+
;
	
List_of_indicators:
	"Indicators:"
	(indicator+=Indicator)+
;

List_of_surveys:
	"Surveys:"
	(survey+=Survey)+
;


List_of_validation_rules:
	"Validation_rules:"
	(validationRule+=ValidationRule)+
;

Topic:
	"topic_id:" name=ID
	"Name:" STRING 
	"Description:" STRING
	("Parent_topic:"	linkParentTopic=[Topic])?
    //  Constraint: avoid cycles	
;

Indicator:
	"Indicator_id:" name=ID
	"Name:" STRING
	"Description:" STRING
	("PreUnit:" STRING)?
	("PostUnit:" STRING)?
	"Topic:" linkTopic=[Topic]
	"Indicator_type": indicator_type=Indicator_type
	"DataType:" datatype=Datatype
    // Constraint: only direct indicators can have datatype list
    "LowerThreshold:" DOUBLE
    "UpperThreshold:" DOUBLE
;

Datatype:
	text="text" | integer="integer" | double="double" | 
	date="date" | boolean="boolean" | singleChoice=SingleChoice 
	| multipleChoice=MultipleChoice
;

MultipleChoice:
	"multipleChoice"
	"List_items:"
	(list_item+=List_item)+
;

SingleChoice:
	"singleChoice"
	"List_items:"
	(listItem+=List_item)+
;

List_item:
	"Order:" INT
	"Text:" STRING
;

Indicator_type:
	direct=Direct | indirect=Indirect
;

Direct:
	direct="Direct"
	("Condition:" expression=Expression)?
 	// Constraint: We should be able to reference answer options
;

Indirect:
	indirect="Indirect"
	"Formula:" formula=Formula
	'Type:' indicatorClassification=INDICATORCLASSIFICATION
;

Formula:
	statement=Statement
;

UnaryNumeric:
	unary_numeric_function=UNARY_NUMERIC_FUNCTION
	"(" expression=Expression ")"
;

BinaryNumeric:
	binary_numeric_function=BINARY_NUMERIC_FUNCTION
	"(" expressionl=Expression "," (INT | STRING) ")"
;

Statistical:
	statistical_function=STATISTICAL_FUNCTION
	"(" "["referenceIndicator=[Indicator]"]" ")"
 	//Constraint: only direct indicators should be used
;

Statement:
	expression=Expression | if_statement=If_statement
;

If_statement:
	"IF" expression=Expression "THEN" thenStatement=Statement
	(=>"ELSE" elseStatement=Statement)?
;

Expression:
	simpleExpressionl=Simple_expression ( ("=" | "<>" | "<" | "<=" | ">" | ">=" | "==") 
	simpleExpressionr=Simple_expression)?
;

Simple_expression:
	terml=Term ((("+" | "-") | "OR") termr=Term)*
;

Term:
	factorl=Factor ((("*" | "/") | "AND") factorr=Factor)*
;

Factor:
	basel=Base ("^" baser=Base)?
;

Base:
	( "(" expression=Expression ")" | "["referenceIndicator=[Indicator] "]" | 
	statistic=Statistical | unarynumeric=UnaryNumeric |
	binarynumeric=BinaryNumeric | BOOLEAN | STRING | INT | DOUBLE )
    // We should add a constraint so that statistical functions cannot be used in conditions of validation rules
    // In formulas indicators should not reference themselves
;

Survey:
	"survey_id:" name=ID
	"Name:" STRING
	"Description:" STRING
	"SurveyType:" surveytype+=SURVEYTYPE
	("WelcomeTxt:" STRING)?
	("ClosingTxt:" STRING)?
	"MinThreshold:" DOUBLE
	("Anonymous:" BOOLEAN)?
	list_of_sections+=List_of_sections
;

List_of_sections:
	"Sections:"
	(section+=Section)+
;

Section:
	"section_id:" name=ID
	"Title:" STRING
	"Order:" INT
    ("Topic:" linkTopic=[Topic])?
    (list_of_subsections+=List_of_subsections)+
;

List_of_subsections:
	"Subsections:"
	(subsection+=Subsection)+
;

Subsection:
	"subsection_id:" name=ID
	"Title:" STRING
    ("Topic:" linkTopic=[Topic])?
    (list_of_questions+=List_of_questions)
	(list_of_fragments+=List_of_fragments)?
;

List_of_fragments:
	"TextFragments:"
	(text_fragment+=Text_fragment)*
;

Text_fragment:
	"Text:" STRING
	"Order:" INT
;

List_of_questions:
	"Questions:"
	(question+=Question)+
;

Question:
	"question_id:" name=ID
	"Name:" STRING
	"Description:" STRING
	"isMandatory:" BOOLEAN
	"UIComponent:" uicomponent+=UICOMPONENT
	"Order:" INT
	"Indicator:" linkIndicator=[Indicator]?
 	// Constraint: questions can only be linked to direct indicators
	"Instruction:" STRING
;

List_of_certification_levels:
	"Certification_levels:"
	(certification_level+=Certification_level)*
;

Certification_level:
	"certification_id:" name=ID
	"Name:" STRING
	"Description:" STRING
	"Level:" DOUBLE 
	"Colour:" STRING
	list_of_requirements=List_of_requirements
;

List_of_requirements:
	"Requirements:"
	"["referenceRequirement+=[Indicator]"]"("
	,""["referenceRequirement+=[Indicator]"]")* 
;

ValidationRule:
	"Type:" ruletype=RULETYPE
	"Condition:" expression=Expression
	"Message:" STRING
;
// So far, the validation rules should only be triggered when the user validates the data or submits the survey response. It should only be possible to submit when there are no errors. And when there are warnings someone should confirm that they want to submit the data with the warnings

enum INDICATORCLASSIFICATION: performance="performance" | score="scoring" | certification="certification";
enum RULETYPE: warning="warning" | error="error" ;
enum UNARY_NUMERIC_FUNCTION: absolute="abs" | int="int" ;
//abs explanation: https://support.google.com/docs/answer/3093459 , int explanation: https://support.google.com/docs/answer/3093490
enum BINARY_NUMERIC_FUNCTION: roundup="roundUp" |
rounddown="roundDown" | round="round" 
| countif="countIf" ; 
// Round up explanation: https://support.google.com/docs/answer/3093443 , round down: https://support.google.com/docs/answer/3093442 , round: https://support.google.com/docs/answer/3093440, countIf: https://support.google.com/docs/answer/3093480?hl=en-GB
enum UICOMPONENT: field="field" | line="line" | 
textBox="textBox" | checkBox="checkBox" | 
dropDown="dropDown" | radioButton="radioButton";
enum STATISTICAL_FUNCTION: minimum="min" 
| maximum="max" | sum="sum" | mean="avg" 
| mode="mode" | median="median";
enum SURVEYTYPE: multi="multi" | single="single" ;
terminal BOOLEAN : ("true"|"false");
terminal DOUBLE: INT "." INT;
\end{lstlisting}

\section{Conclusion}
Currently, we are working on a third version of the metamodel and grammar. We are extending the metamodel, so that it includes necessary classes to perform ESGA audits. Moreover, we wish extend the metamodel and grammar, so more types of impact measurement methods can supported with openESEA. Meaning, in the future we want to be able to create models for social impact assessment methods, life-cycle assessment methods, as well as other families of methods. openESEA should be able to interpret models created with each new version of the grammar, therefore openESEA evolves as the metamodel and grammar evolve.

\section*{Acknowledgements}
We appreciate the contributions made by many students from the Master in Business Informatics at Utrecht University while conducting their master theses within the openESEA umbrella project. We are thankful to all the experts in ESGA methods who shared their time and knowledge with us so we could analyse and compare many ESGA methods; we later generalised those cases and engineered the openESEA domain specific language so it was versatile enough to allow specifying such methods. We have formatted the Xtext grammar using the styles defined in \url{https://gist.github.com/wildangerm/d7c6b7edaf2aaf45d07336e1200df5a9}. Sergio España is supported by a María Zambrano grant of the Spanish Ministry of Universities, co-funded by funds of the Next Generation EU European Recovery Plan.

\bibliographystyle{unsrtnat}
\bibliography{references} 

\end{document}